\documentclass{SCGE}
\usepackage[colorlinks,linkcolor=blue,citecolor=blue,urlcolor=blue]{hyperref}
\usepackage{graphicx}
\usepackage{amsmath}
\usepackage{amsfonts}
\usepackage{amssymb}
\usepackage{mathrsfs}
\usepackage{latexsym}
\usepackage{dsfont}
\usepackage{graphicx}

\newcommand{\tr}{{\rm Tr}}

\renewcommand{\geq}{\geqslant}
\renewcommand{\leq}{\leqslant}

\newcommand{\trace}{\mathop{\rm Tr}\nolimits}
\def\be{\begin{equation}}
\def\ee{\end{equation}}
\def\ba{\begin{array}}
\def\ea{\end{array}}

\def\qed{\leavevmode\unskip\penalty9999 \hbox{}\nobreak\hfill
     \quad\hbox{\leavevmode  \hbox to.77778em{%
               \hfil\vrule   \vbox to.675em%
               {\hrule width.6em\vfil\hrule}\vrule\hfil}}
     \par\vskip3pt}
\newcommand{\dl}{\delta }

\let\citedash\relax
\makeatletter \providecommand{\citedash}{\hbox{-}\penalty\@m}
\makeatother

\begin{document}
%%%%%%新版式要加上这组
%\begin{picture}(0,0){\rm
%\put(0,-20){\makebox[160truemm][l]{\bf {\sanhao\raisebox{2pt}{.}}
%Article  {\sanhao\raisebox{1.5pt}{.}}}}}
%\put(0,-34){\jiuwuhao {\textcolor[rgb]{0.5,0.5,0.5}{\sf %Special Topic: Fluid Mechanics
%}}}%%(11月注释：调\textcolor[rgb]{x,x,x}中的数字x越大越灰)
%\end{picture}

%\input psfig.sty
\def\bm{\boldsymbol}

\def\dl{\displaystyle}
\def\du{\end{document}}
\def\d{{\rm d}}
\def\e{{\rm e}}
\def\i{{\rm i}}

% The author doesn't need fill in it.
%\Year{2016} %
%\Month{January}
%\Vol{59} %  卷号
%\No{1} %  期号
\BeginPage{1} % 起页码
%\AuthorMark{{\rm A. Author}, et al.}  %(11月注释：页眉上的作者)
%\AuthorMarkCite{A. Author, B. Author, C. Author, and D. Author} %(11月注释：citation中的作者)
%\DOI{} % The author doesn't need fill in it.
%\ArtNo{000000}

% \title[short text for running head]{full title}{comments for title}
\title[\xiaosihao Sharp Continuity Bounds for Entropy and Conditional Entropy]{Sharp Continuity Bounds for Entropy and Conditional Entropy}%标题 {有黑体的时候需要将标题复制在中括号里面，使得引用条显示白体。没有黑体的时候中括号可以删掉}
\author[1]{Zhihua Chen}{}
\author[2\dag]{Zhihao Ma}{}
\author[3]{Ismail Nikoufar}{}
\author[4,5*]{Shao-Ming Fei}{}
\footnote{Corresponding author (email: \dag mazhihaoquantum@126.com, *feishm@cnu.edu.cn)}

\address[{\rm1}]{Department of Applied Mathematics,  Zhejiang
University of Technology, Hangzhou, 310014, China;}
\address[{\rm2}]{School of Mathematical Sciences, Shanghai Jiaotong
University, Shanghai, 200240, China;}
\address[{\rm3}]{Department of Mathematics, Payame Noor University, 19395-3697 Tehran, Iran;}
\address[{\rm4}]{School of Mathematical Sciences, Capital Normal University, Beijing 100048, China\\
\rm5 Max-Planck-Institute for Mathematics in the Sciences, 04103 Leipzig, Germany}

\maketitle

%\vspace{-3.5mm}
%{\footnotesize\begin{center} Received January 1, 2016; accepted January 1, 2016; published online January 1, 2016%收稿日期
%\end{center}}
%\vspace*{-5mm}

%     Abstract is required.
%\begin{center}
%\rule{16.5cm}{0.4pt}
%\parbox{16.5cm}
%{\begin{abstract}
%Besides von Neumann entropy, R$\acute{e}$nyi entropy and Tsallis entropy are also two
%important entropies and play an essential role in quantum information theory.
%We study the continuity estimation of the R$\acute{e}$nyi entropy and the special conditional Tsallis entropy, i.e. conditional linear entropy.
%An inequality relating the R$\acute{e}$nyi entropy difference of two quantum states to their
%trace norm distance is derived.
%This inequality is shown to be tight in the sense that equality can be attained for
%every prescribed value of the trace norm distance.
%It includes the sharp Fannes inequality for von Neumann entropy as a special case.
%We also get the continuity of linear entropy discord from the continuity of conditional linear entropy. %摘要
%\end{abstract}}
%\end{center}\vspace*{-0.6cm}

%\begin{center}
%\parbox{16.5cm}
%{\bf\jiuhao Fannes inequality, R$\acute{e}$nyi entropy, Continuity estimation}%关键词
%\end{center}

%\begin{center}
%{\PACS{\rm 03.65.Ud, 03.67.Mn}}%分类号(3~5 codes)  http://phys.scichina.com:8083/sciGe/UserFiles/File/pacs.pdf
%\CITA    %%(11月注释：Citation内容自动生成)
%\Cit{~~~???, et al. ???. Sci China-Phys Mech Astron, 2014, 57: 1--6, doi:}%%(11月注释：Citation内容需手动填写)
%\end{center}

\textwidth=178truemm \textheight=236truemm%%%%%%新版式要加上

%%%%%%%%%%%%%%%%%%%%%%%%%%%%%%%%%%%%%%%%%%%%%%%%%%%%%%%%%%%%
%\wuhao\vspace*{1.5mm}

\begin{multicols}{2}

%%%%%%%%%%%%%%%%%%%%%%%%%%%%%%%%%%%%%%%%%%%%%%%%%%%%%%%%%%%%
%% Text of article.
%%%%%%%%%%%%%%%%%%%%%%%%%%%%%%%%%%%%%%%%%%%%%%%%%%%%%%%%%%%%
%    Section headings
\renewcommand{\baselinestretch}{1.08} \baselineskip 12.2pt\parindent=10.8pt

\renewcommand{\thefootnote}

\noindent Dear Editor,

The von Neumann entropy captures many operational
quantities in the quantum information theory such as quantum capacity of the communication channel. Von Neumann entropy is continuous and
is represented by Fannes inequality, which was originally given in\cite{Fannes}. Quantum correlations such as entanglement and quantum discord
, et al., are important resources in quantum information processing. In the last year enormous progress on the generation, concentration, detection and quantification of entanglement has been achieved \cite{Mai}.
Fannes inequality has many
applications in the quantum information theory, such as the investigation of continuity of entanglement measures, including entanglement of formation, relative entropy of entanglement, squashed entanglement and conditional entanglement of mutual information, and the continuity of quantum channel capacities \cite{D.Y}.
Recently Fannes inequality was improved to get a sharp one and it was also generalized to Tsallis entropy \cite{Audenaert,A.E}.

However, in non-asymptotic  settings, the natural quantities that arise are R$\acute{e}$nyi entropies \cite{0} and
the properties of R$\acute{e}$nyi entropies were also investigated in many papers, such as \cite{Lin}.
R$\acute{e}$nyi entropies have many applications, as in the case of one-shot problems, typically arising in cryptographic
settings, the min- and max-entropies are widely used \cite{0}.
In \cite{Adesso}, the authors found that R$\acute{e}$nyi-2 entropy was a proper measure of information for any multimode Gaussian
state, and they defined and analyzed
the measures of Gaussian entanglement and quantum correlation by using R$\acute{e}$nyi-2 entropy, and found its  properties such as monogamy. In our work, we study the continuity property of R$\acute{e}$nyi-$\alpha$ entropy, which includes R$\acute{e}$nyi-2 entropy as a special case. Our result is also useful in studying the continuity of the entanglement measure of the Gaussian state in quantum harmonic systems.

On the other hand, the authors found that Tsalli-2 entropy (i.e., linear entropy) was natural to define the measure  of quantum correlation for the the discrete system
\cite{Ma15}. They called this measure  as linear discord and used conditional linear entropy to define the linear discord. They found that the linear discord has deep connection with the original discord defined by von Neumann entropy. Moreover, they gave the analytical formula for arbitrary $2\otimes d$ state of the linear discord. However, a question still remains open: if two states are close, is their linear discord also close to each other?
In other words, is the linear discord continuous?
For the original discord, the answer is affirmative, see \cite{Xi}. For the linear discord, there is no answer yet. Hence it is worthwhile to study the continuity of conditional linear entropy.

We have two aims in this work: first, we study the continuity estimation of the R$\acute{e}$nyi entropy and present
a tight inequality relating the R$\acute{e}$nyi entropy difference of two quantum states to their
trace norm distance, which includes the sharp Fannes inequality for von Neumann entropy as a special case. Second, we study the continuity of conditional linear entropy, and prove a useful property for a measure of the quantum correlation: linear discord.

The von Neumann entropy of a quantum state $\rho$ is defined by
\begin{equation}
S(\rho):=-\trace[\rho\log_2\rho].
\label{vN}
\end{equation}
For the classical probability distributions, the von Neumann entropy reduces to the Shannon entropy,
\begin{equation}
H(p):=\sum_i^d H(p_i)= -\sum_i^d p_i\log_2 p_i,
\end{equation}
where $p:=(p_i)=(p_1,p_2,...,p_d)$ is a $d$-dimensional probability vector, $p_i\geq 0$, $\sum_i^d p_i=1$ and
$H(p_i):= - p_i\log_2 p_i$.

In \cite{Fannes} Fannes proved his famous inequality for the continuity of the von Neumann entropy,
\begin{equation}\label{Fannes}
|S(\rho)-S(\sigma)| \leq 2T\log_2(d)-2T\log_2(2T),
\end{equation}
where $T:=\frac{||\rho-\sigma||_{1}}{2}$ is half of the trace norm distance between the states
$\rho$ and $\sigma$, $||\rho-\sigma||_{1}=\tr[|\rho-\sigma|]$, and
$|X|:=\sqrt{(X)^\dag(X)}$ denotes the absolute value of an operator $X$. Obviously $T\in [0,1]$. The inequality (\ref{Fannes}) is valid for $0\leq T\leq 1/2e$, where $e$ is Euler's number.
The inequality (\ref{Fannes}) is further improved to be a sharp one by Audenaert \cite{Audenaert}:
\begin{equation}
|S(\rho)-S(\sigma)| \leq T\log_2(d-1)-H((T,1-T)).
\label{Audenaert}
\end{equation}

The R$\acute{e}$nyi entropy is a more general form of the von Neumann entropy,
\be\label{Tsallis}
 H_{\alpha}(\rho):=\frac{1}{1-\alpha}\log [\tr(\rho^{\alpha})],~~~~ \alpha> 0,
\ee
when $\alpha$ goes to one, R$\acute{e}$nyi entropy becomes the von Neumann entropy.
In the following we show that for the R$\acute{e}$nyi  entropy, an improved sharp Fannes-type inequality exists.

{\bf Theorem 1.} For all $d$-dimensional quantum states $\rho$ and $\sigma$,
\begin{equation}\label{ma-1}
|H_{\alpha}(\rho)-H_{\alpha}(\sigma)|\leq \frac{d^{\alpha-1}}{1-\alpha}[1-(1-T)^{\alpha}-(d-1)^{1-\alpha}T^{\alpha}],~~ \alpha> 1,
\end{equation}
\begin{equation}\label{ma-1b}
|H_{\alpha}(\rho)-H_{\alpha}(\sigma)|\leq \frac{1}{1-\alpha}[1-(1-T)^{\alpha}-(d-1)^{1-\alpha}T^{\alpha}],~~ \alpha< 1,
\end{equation}
where $T$ is the trace norm distance of $\rho$ and $\sigma$. See proof in Appendix.

We investigated the continuity estimation of the R$\acute{e}$nyi entropy, by
presenting an inequality which relates the R$\acute{e}$nyi entropy difference of two quantum states to their
trace norm distance. In our inequality, equality can be attained for
every prescribed value of the trace norm distance. It is direct to verify that for
$\alpha\to 1$, our inequality (\ref{ma-1}) and (\ref{ma-1b}) give rise to
the sharp Fannes inequality for von Neumann entropy.
It has potential applications in investigating the continuity of entanglement measure and more general correlations for multimode Gaussian
states, since R$\acute{e}$nyi-2 entropy is a proper information measure for this kind of state \cite{Adesso}.

Besides R$\acute{e}$nyi entropy, linear entropy is also used to measure quantum correlations, such as linear discord \cite{Ma15}, which is defined as the minimal  difference of the two  conditional linear entropy, before and after the local projective measurement, $D_{2}(\rho_{AB}):=\min\limits_{P_{i}} (S_{2}(A|B)-S_{2}(P_{i}|B))$, where $S_{2}(A|B)$ is the conditional linear entropy of the original state $\rho_{AB}$, while $S_{2}(P_{i}|B)$ is the conditional linear entropy of the post measurement state after local measurement $P_{i}$, and the minimum runs over all local projection measurements $P_{i}$. Therefore it is also
important to study the continuity of the linear entropy, especially conditional linear entropy.
It can help us get the continuity of the linear discord.

We can prove the following conclusion: conditional linear entropy is continuous, see proof in Appendix.

{\bf Theorem 2.} For bipartite quantum states $\rho_{AB}$ and $\sigma_{AB}$, if $\epsilon:=||\rho_{AB}-\sigma_{AB}||_{1}< 1$,
 then the following inequality holds,
\begin{equation}\label{ma-5}
|S_{2}(\rho_{AB}|\rho_{B})-S_{2}(\sigma_{AB}|\sigma_{B})|\leq 4\epsilon +2h_{2}(\epsilon,1-\epsilon).
\end{equation}

By using the method of \cite{Xi}, it is straightforward to show that the linear discord  $D_{2}(\rho_{AB})$ is also continuous:

{\bf Theorem 3.} For bipartite quantum states $\rho_{AB}$ and $\sigma_{AB}$, if $\epsilon:=||\rho_{AB}-\sigma_{AB}||_{1}< 1$,
 then
\begin{equation}\label{ma-5-1}
|D_{2}(\rho_{AB})-D_{2}(\sigma_{AB})|\leq 8\epsilon +4h_{2}(\epsilon,1-\epsilon).
\end{equation}

In summary, we have investigated the continuity of  R$\acute{e}$nyi entropy and conditional linear entropy.
Through the continuity of conditional linear entropy, the continuity of linear discord has also been obtained, which
means that the linear discord varies as the quantum state changes continuously.
This fact can guarantee that the errors in state tomograph would not significantly affect the result of the quantum correlations in the state.
As the sharp Fannes inequality is the special case of our theorem about the continuity of R$\acute{e}$nyi entropy, our results can also be used to
verify the continuity of entanglement measures for continuous variable quantum states.

\bigskip
\noindent{\bf Acknowledgments}\, \, This work is supported by the
NSFC under number 11371247, 11275131, 11675113 and 11571313.

\newpage
\begin{appendix}
\setcounter{section}{0}
\def\thesection{Appendix}

{\bf{Appendix}}

{\bf{Proof of Theorem 1:}}

{\bf{Lemma 1.1.}} For all $d$-dimensional quantum states $\rho$ and $\sigma$,
\begin{equation}\label{ma-2}
|H_{\alpha}(\rho)-H_{\alpha}(\sigma)|\leq  d^{\alpha-1}|S_{\alpha}(\rho)-S_{\alpha}(\sigma)|~~~~ \alpha> 1,
\end{equation}
\begin{equation}\label{ma-2b}
|H_{\alpha}(\rho)-H_{\alpha}(\sigma)|\leq  |S_{\alpha}(\rho)-S_{\alpha}(\sigma)|~~~~ \alpha< 1.
\end{equation}
where $S_{\alpha}(\rho):=[\tr(\rho^{\alpha})-1]$.

The above inequalities are obtained by using Cauchy mean value theorem.

%Hence in stead of (\ref{ma-1} and \ref{ma-1b}), we prove the
%following inequality,

{\bf{Lemma 1.2.}} For all $d$-dimensional quantum states $\rho$ and $\sigma$,
\begin{equation}\label{ma-1p}
|S_{\alpha}(\rho)-S_{\alpha}(\sigma)|\leq 1-(1-T)^{\alpha}-(d-1)^{1-\alpha}T^{\alpha}.
\end{equation}

{\bf{Lemma 1.3.}} For all probability distributions $p=(p_{i})$ and $q=(q_{i})$, the following inequality holds:
\begin{equation}\label{ma-1pp}
|S_{\alpha}(p)-S_{\alpha}(q)|\leq 1-(1-T)^{\alpha}-(d-1)^{1-\alpha}T^{\alpha},
\end{equation}
where $T=\frac{1}{2}\sum\limits_{i=1}^{d}|p_i-q_i|$, $S_{\alpha}(p)=\sum\limits_{i}[(p_{i})^{\alpha}-p_{i}]
=\sum\limits_{i}S_{\alpha}(p_{i})$, $S_{\alpha}(q)=\sum\limits_{i}[(q_{i})^{\alpha}-q_{i}]
=\sum\limits_{i}S_{\alpha}(q_{i})$, $p_i\geq 0,$ $q_i\geq 0 $ and $\sum\limits_i p_i=\sum\limits_i q_i=1$, $S_{\alpha}(p_{i}):=[(p_{i})^{\alpha}-p_{i}]$.

Let $\lambda_{i}$, $i=1,2,...,d$, be the eigenvalues of $\rho,$
one has $S_{\alpha}(\rho)=\sum\limits_{i}[(\lambda_{i})^{\alpha}-\lambda_{i}]:=\sum\limits_{i}S_{\alpha}(\lambda_{i})$, where  $S_{\alpha}(\lambda_{i}):=[(\lambda_{i})^{\alpha}-\lambda_{i}]$.

Firstly, we prove Lemma 1.3. Then Lemma 1.2 is also proved for the diagonal quantum states $\rho$ and $\sigma.$

{\bf{Proof of Lemma 1.3:}} Let $q=p+\delta^{+}-\delta^{-}$,
where $\delta^{+}=(\delta^{+}_i)$ and $\delta^{-}=(\delta^{-}_i)$ are two vectors such that
$\delta^{+}_i\geq 0$, $\delta^{-}_i\geq 0$, $i=1,2,...,d$,
$\delta^{+}\cdot\delta^{-}=0$, $\sum_i\delta_i^{+}=T$.
We prove (\ref{ma-1pp}) of Lemma 1.3 in three cases according to the values of $\alpha:$
$\alpha<1,$ $1\leq\alpha<2$ and $\alpha\geq 2.$

({\bf{Case I}}) $\alpha<1:$

{\bf{(a).}} $S_{\alpha}(p)$ is concave, $S_{\alpha}(p+\delta^{+}-\delta^{-})-S_{\alpha}(p)$
is a concave  function with respect to $\delta^{+}$. It gets its minimum
at a certain point, say, $\delta^{+}=e^1$.

Then $T=e^1,$  $p=(p_1,(1-p_1)r)$, $q=(p_1+T,(1-p_1)r-T s)$, where $r$ and $s$ are $d-1$ dimensional probability vectors such that
$p_1+T\leq 1$, $(1-p_1)r-T s\geq 0$, and $T s=\delta^{-}$. We have
\begin{eqnarray}
\nonumber
&&S_{\alpha}(q)-S_{\alpha}(p)=S_{\alpha}(p_1+T)-S_{\alpha}(p_1)\\ \nonumber
&&~~~~~~~~~~~~~~~~~~~~~~~~~~~+S_{\alpha}((1-p_1)r-Ts)-S_{\alpha}((1-p_1)r).
\end{eqnarray}

{\bf{(b).}} Denote $(1-p_1)r-Ts=(1-p_1-T)\eta$, then
\begin{eqnarray}\label{p1}
&&S_{\alpha}((1-p_1)r-Ts)-S_{\alpha}((1-p_1)r)\\ \nonumber
&&=S_{\alpha}((1-p_1-T)\eta)-S_{\alpha}((1-p_1-T)\eta+Ts).
\end{eqnarray}

Since $S_{\alpha}(x)-S_{\alpha}(x+y)$ is concave and a monotonously increasing function of $x$, the right-hand side of (\ref{p1})
gets its minimum at certain point of $\eta$, say, $\eta=e^1.$
Let $s=(s_1,(1-s_1)\phi)$ with $\phi$ a $d-2$ dimensional probability vector,
we have
\begin{eqnarray}
\nonumber
&&S_{\alpha}((1-p_1-T)\eta)-S_{\alpha}((1-p_1-T)\eta+Ts)\\ \nonumber
&&=S_{\alpha}(1-p_1-T)-S_{\alpha}(1-p_1-T(1-s_1))\\ \nonumber
&&~~~-S_{\alpha}(T(1-s_1)\phi)\\ \nonumber
&&\triangleq \Delta.
\end{eqnarray}
As $S_{\alpha}(T(1-s_1)\phi)$ gets its maximum when $\phi$ is the uniform distribution $=(1,1,\cdots,1)/(d-2)$,
$\Delta$ has the minimum,
\begin{eqnarray}
\nonumber
&&\Delta=S_{\alpha}(1-p_1-T)-S_{\alpha}(1-p_1-T(1-s_1))\\ \nonumber
&&~~~-(d-2)^{1-\alpha}S_{\alpha}(T(1-s_1)).
\end{eqnarray}

{\bf{(c).}} From
$$
\frac{\partial \Delta}{\partial s_1}=-T \alpha[(1-p_1-T(1-s_1))^{\alpha-1}-(d-2)^{1-\alpha}(T(1-s_1))^{1-\alpha}]=0,
$$
we get
\be\label{p2}
T(1-s_1)=\frac{(1-p_1)(d-2)}{(d-1)}\equiv \omega.
\ee

When $0<T<\omega$, there is no local minimum of $\Delta$ from (\ref{p2}).
For $T(1-s_1)=T$, i.e. $s_1=0$, $\Delta$ has a minimum $-(d-2)^{(1-\alpha)}S_{\alpha}(T)$.
Therefore $S_{\alpha}(q)-S_{\alpha}(p)$ gets its minimum
$S_{\alpha}(p_1+T)-S_{\alpha}(p_1)-(d-2)^{1-\alpha}S_{\alpha}(T)$. Moreover, due to that $S_{\alpha}(p_1+T)-S_{\alpha}(p_1)$
is a decreasing function of $p_1$, when $p_1=1-\frac{(d-1)T}{d-2}$, $S_{\alpha}(q)-S_{\alpha}(p)$ gets its minimum
$(1-\frac{T}{d-2})^{\alpha}-(1-\frac{(d-1)T}{d-2})^{\alpha}-(d-2)^{1-\alpha}T^{\alpha}$.

When $\omega\leq T\leq 1-p_1$, (\ref{p2}) can be satisfied and
$\Delta$ gets its minimum $(1-p_1-T)^{\alpha}-{(1-p_1)^{\alpha}}/{(d-1)^{\alpha-1}}$.
Therefore $S_{\alpha}(q)-S_{\alpha}(p)$ gets its minimum
$$
-S_{\alpha}(p_1)+S_{\alpha}(p_1+T)+(1-p_1-T)^{\alpha}-\frac{(1-p_1)^{\alpha}}{(d-1)^{\alpha-1}}.
$$
The derivative of the above formula with respect to $p_1$ is less than zero.
Hence, when $p_1=1-T$, $S_{\alpha}(q)-S_{\alpha}(p)$ gets its minimum $1-(1-T)^{\alpha}-(d-1)^{1-\alpha}T^{\alpha}$.

Since $S_{\alpha}(x)-S_{\alpha}(x-T)$ is a decreasing function of $x$ and $1>1-\frac{T}{d-2}$,
$$
1-(1-T)^{\alpha}\leq(1-\frac{T}{d-2})^{\alpha}-(1-\frac{(d-1)T}{d-2})^{\alpha}.
$$
Therefore
$1-(1-T)^{\alpha}-(d-1)^{1-\alpha}T^{\alpha}\leq (1-\frac{T}{d-2})^{\alpha}-(1-\frac{(d-1)T}{d-2})^{\alpha}-(d-2)^{1-\alpha}T^{\alpha}$ and then
$1-(1-T)^{\alpha}-(d-1)^{1-\alpha}T^{\alpha}$ is the minimum of $S_{\alpha}(q)-S_{\alpha}(p) $.

({\bf{Case II}}) $1\leq \alpha<2:$

{\bf{(a).}} $S_{\alpha}(p)-S_{\alpha}(q)=S_{\alpha}(p)-S_{\alpha}(p+\delta^{+}-\delta^{-})$ is concave with respect to
$\delta^{+}$. Take $\delta^{+}=e^1.$ We have
\begin{eqnarray}
\nonumber
&&S_{\alpha}(p)-S_{\alpha}(q)=S_{\alpha}(p_1)-S_{\alpha}(p_1+T)\\ \nonumber
&&+S_{\alpha}((1-p_1)r)-S_{\alpha}((1-p_1)r-Ts),
\end{eqnarray}
where
\begin{eqnarray}
\nonumber
&&S_{\alpha}((1-p_1)r)-S_{\alpha}((1-p_1)r-Ts)\\ \nonumber
&&=S_{\alpha}((1-p_1-T)\eta+Ts)-S_{\alpha}((1-p_1-T)\eta).
\end{eqnarray}

{\bf{(b).}} $S_{\alpha}(x+y)-S_{\alpha}(x)$ is concave with respect to $x$ for $\alpha<2$, $S_{\alpha}((1-p_1-T)\eta+Ts)-S_{\alpha}((1-p_1-T)\eta)$
gets its minimum at $\eta=e^1$.
Let $s=(s_1,(1-s_1)\phi)$ with $\phi$ a $d-2$ dimensional probability vector. We get
\begin{eqnarray}
\nonumber
&&S_{\alpha}((1-p_1-T)\eta+Ts)-S_{\alpha}((1-p_1-T)\eta)\\ \nonumber
&&=S_{\alpha}(1-p_1-T(1-s_1))+S_{\alpha}(T(1-s_1)\phi)\\ \nonumber
&&-S_{\alpha}(1-p_1-T)\triangleq \Delta.
\end{eqnarray}
When $\phi=(1,1,\cdots,1)/(d-2)$, $S_{\alpha}(T(1-s_1)\phi)$ gets its minimum, and
$\Delta$ gets its minimum, that is $S_{\alpha}(1-p_1-T(1-s_1))-S_{\alpha}(1-p_1-T)+(d-2)^{1-\alpha}S_{\alpha}(T(1-s_1))$.

{\bf{(c)}} From $\frac{\partial \Delta}{\partial s_1}=0$, we have the formula (\ref{p2}) again.

When $0<T<\omega,$ $\Delta$ has no local minimum.
For $T(1-s_1)=T$, $\Delta$ has a minimum $(d-2)^{(1-\alpha)}S_{\alpha}(T)$.
$S_{\alpha}(p)-S_{\alpha}(q)$ gets its minimum
$S_{\alpha}(p_1)-S_{\alpha}(p_1+T)+(d-2)^{1-\alpha}S_{\alpha}(T)$, which takes the minimum
value $(1-\frac{(d-1)T}{d-2})^{\alpha}-(1-\frac{T}{d-2})^{\alpha}+(d-2)^{1-\alpha}T^{\alpha}$
at $p_1=1-\frac{(d-1)T}{d-2}$.

For $\omega\leq T\leq 1-p_1$, at $T(1-s_1)=\omega$, $\Delta$ gets its minimum
$\frac{(1-p_1)^{\alpha}}{(d-1)^{\alpha-1}}-(1-p_1-T)^{\alpha}$. $S_{\alpha}(q)-S_{\alpha}(p)$ gets its minimum
$S_{\alpha}(p_1)-S_{\alpha}(p_1+T)-(1-p_1-T)^{\alpha}+\frac{(1-p_1)^{\alpha}}{(d-1)^{\alpha-1}}
=(1-T)^{\alpha}+(d-1)^{1-\alpha}T^{\alpha}-1$ at $p_1=1-T$. Therefore
\begin{eqnarray}
\nonumber
&&(1-T)^{\alpha}+(d-1)^{1-\alpha}T^{\alpha}-1\\ \nonumber
&&<(1-\frac{(d-1)T}{d-2})^{\alpha}-(1-\frac{T}{d-2})^{\alpha}+(d-2)^{1-\alpha}T^{\alpha}
\end{eqnarray}
and (\ref{ma-1pp}) is valid when $1\leq\alpha<2$.

({\bf{Case III}}) $\alpha\geq 2:$ Following the same step as {\bf{Case II}}, we get that $S_{\alpha}((1-p_1)r)-S_{\alpha}((1-p_1)r-Ts)$ is concave with respect to $s$. Hence $S_{\alpha}((1-p_1)r)-S_{\alpha}((1-p_1)r-Ts)$
gets its minimum in one of the extreme points of $s$, say, $s=e^1$.

Let $r=(r_1,(1-r_1)\phi)$ with $\phi$ a $d-2$ dimensional probability vector. Then
\begin{eqnarray}
\nonumber
&&S_{\alpha}((1-p_1)r)-S_{\alpha}((1-p_1)r-Ts)\\ \nonumber
&&=S_{\alpha}((1-p_1)r_1)-S_{\alpha}((1-p_1)r_1-T)\triangleq \nabla.
\end{eqnarray}
Since
$$
\frac{\partial\nabla}{\partial r_1}=\alpha (1-p_1)^{\alpha}r_1^{\alpha-1}-\alpha(1-p_1)((1-p_1)r_1-T)^{\alpha-1}>0,
$$
when $(1-p_1)r_1=T$, $\nabla$ has the minimum $T^{\alpha}$. Therefore $S_{\alpha}(p)-S_{\alpha}(q)$ takes
its minimum $p_1^{\alpha}-(p_1+T)^{\alpha}+T^{\alpha}$. Because $p_1^{\alpha}-(p_1+T)^{\alpha}+T^{\alpha}$ decreases
with the decrease of $p_1$, $S_{\alpha}(p)-S_{\alpha}(q)$ takes its minimum $T^{\alpha}+(1-T)^{\alpha}-1$
at $p_1=1-T$.

Now we have proved the inequality (\ref{ma-1pp}), i.e. Lemma 1.3, namely the inequality (\ref{ma-1p}) of Lemma 1.2
for the case that both states $\rho$ and $\sigma$ are diagonal ones.
For general $\rho$ and $\sigma$, the inequality can be directly proved by taking into
account the fact that the  R$\acute{e}$nyi entropy is unitary invariant \cite{1}.
By using Lemma 1.1, Theorem 1 is proved.

{\bf{Proof of Theorem 2:}}

First, we define the relative linear entropy $D(\rho||\sigma):=\tr(\rho^{2})- \tr(\rho\sigma)$. The linear entropy is then given by $S_{2}(\rho)=-D(\rho||I)=1-\tr(\rho^{2})$,
while the conditional linear entropy is given by $S_{2}(\rho_{AB}|\rho_{B}):=S_{2}(A|B):=-D(\rho_{AB}||I\otimes \rho_{B})=\tr(\rho^{2}_{B})-\tr(\rho^{2}_{AB})$.

{\bf Definition } We define the Tsalli $\alpha$ relative entropy as:
\begin{equation}
\hat{T}_{\alpha}(\rho||\sigma)=\frac{1}{\alpha-1}(\tr(\rho^{\alpha})- \tr(\rho\sigma^{\alpha-1})),\label{t3}
\end{equation}
here $\alpha\in (0,\infty)$.
When $\alpha$ goes to one, the Tsalli $\alpha$ relative entropy becomes the (quantum) relative entropy,
$S(\rho||\sigma)=-\tr(\rho\log\sigma)-S(\rho)$.

{\bf Lemma 2.} The relative entropy defined by (\ref{t3})
is jointly convex for $\alpha\in [0,1)\cup(1,2]$.

{\bf Proof.}
The function $f(t)=\frac{1}{\alpha-1}(1-t^{\alpha-1})$ is a convex function for $\alpha\in [0,1)\cup(1,2]$ when $t>0$,
then
\begin{align*}
g(L_{\rho},R_{\sigma})&=L_{\rho}^{1/2}f(L_{\rho}^{-1/2}R_{\sigma} L_{\rho}^{-1/2})L_{\rho}^{1/2}\\
&=\frac{1}{\alpha-1}(L_{\rho}-L_{\rho}^{2-\alpha}R_{\sigma}^{\alpha-1})
\end{align*}
is jointly convex (cf. \cite{Nikoufar2}). It follows that
$$
(\rho, \sigma)\longmapsto \langle g(L_{\rho},R_{\sigma})(X),X \rangle=\tr(X^*g(L_{\rho},R_{\sigma})(X))
$$
is also jointly convex on $\rho, \sigma$, where $X$ is an arbitrary
operator and $\langle\cdot,\cdot\rangle$ is the Hilbert-Schmidt inner product. Taking $X=\rho^{\frac{\alpha-1}{2}}$ we have
\begin{align*}
\langle g(L_{\rho}&,R_{\sigma})(\rho^{\frac{\alpha-1}{2}}),\rho^{\frac{\alpha-1}{2}} \rangle\\
&=\frac{1}{\alpha-1}\tr(\rho^{\frac{\alpha-1}{2}}(L_{\rho}-L_{\rho}^{2-\alpha}R_{\sigma}^{\alpha-1})(\rho^{\frac{\alpha-1}{2}}))\\
&=\hat{T}_{\alpha}(\rho,\sigma).
\end{align*}
This completes the proof.

{\bf Corollary 1.} The linear entropy  $S_{2}(\rho)$ and the conditional linear entropy $S_{2}(A|B)$ are concave.

{\bf Lemma 3.} (Projective measurements increase entropy) Suppose that $P_{i}$ is a
complete set of orthogonal projectors and $\rho$ is a density operator. Then the
linear entropy of the state $\rho^{'}:=\sum\limits_{i}P_{i}\rho P_{i}$ after the measurement will not decrease, that is, $S_{2}(\rho^{'})\geq S_{2}(\rho)$.

{\bf Proof.} We know that the square of Hilbert-Schimidt metric is defined by $D^{2}_{2}(\rho^{'}, \rho):=\tr(\rho^{'}-\rho)^{2}$, which is nonnegative.
We have $\tr(\rho^{'}-\rho)^{2}=S_{2}(\rho^{'})- S_{2}(\rho)\geq 0$.

{\bf Lemma 4.}  Suppose $\rho=\sum\limits_{i}p_{i}\rho_{i}$, where $\{p_{i}\}$ are some set of probabilities and
$\{\rho_{i}\}$ are density operators. Then we have the following upper bound: $S_{2}(\rho)\leq h_{2}(p_{i})+\sum\limits_{i}p_{i}S_{2}(\rho_{i})$, where $h_{2}(p_{i}):=1-\sum\limits_{i}p^{2}_{i}$.

{\bf Proof.} The proof uses the method similar to the von Neumann entropy case (see Theorem 11.10 of \cite{nielsen}). From direct calculation and the Cauchy- Schwartz inequality, we get the result.

In the following text, we consider the bipartite quantum states on $H\otimes H$, with $d$ being the dimension of Hilbert space $H$.

{\bf Lemma 5.} For the bipartite quantum state $\rho_{AB}$, the following inequality holds,
\begin{equation}\label{ma-3}
|S_{2}(A|B)|\leq \frac{d-1}{d}.
\end{equation}

{\bf Proof.} $S_{2}(A|B)\leq \frac{d-1}{d}$ comes from the subadditivity of the Tsallis entropy (see \cite{Audenaert2}), $S_{2}(\rho_{AB})\leq S_{2}(\rho_{A})+S_{2}(\rho_{B})$. On the other hand,
 we have $S_{2}(A|B)=S_{2}(\rho_{AB})-S_{2}(\rho_{B})\geq -S_{2}(\rho_{B})\geq -\frac{d-1}{d}$, which completes the proof.

{\bf Lemma 6.} For bipartite quantum states $\rho_{AB}$ and $\tau_{AB}$, assume that $0\leq \epsilon\leq 1$, define $\gamma_{AB}:=(1-\epsilon)\rho_{AB}+\epsilon \tau_{AB}$,
 then
\begin{equation}\label{ma-4}
|S_{2}(\rho_{AB}|\rho_{B})-S_{2}(\gamma_{AB}|\gamma_{B})|\leq 2\epsilon\frac{d-1}{d} +h_{2}(\epsilon,1-\epsilon).
\end{equation}

{\bf Proof.} The proof is similar to that of \cite{Fannes}. First, from the concavity of the entropy, we have
$S_{2}(\gamma_{B})\geq (1-\epsilon)S_{2}(\rho_{B})+\epsilon S_{2}(\tau_{B})$. From the upper bound  $S_{2}(\gamma_{AB})\leq h_{2}(\epsilon,1-\epsilon)+(1-\epsilon)S_{2}(\rho_{AB})+\epsilon S_{2}(\tau_{AB})$, we get $S_{2}(\gamma_{AB}|\gamma_{B})=S_{2}(\gamma_{AB})-S_{2}(\gamma_{B})\leq h_{2}(\epsilon,1-\epsilon)+(1-\epsilon)S_{2}(\rho_{AB}|\rho_{B})+\epsilon S_{2}(\tau_{AB}|\tau_{B})$. Therefore $S_{2}(\rho_{AB}|\rho_{B})-S_{2}(\gamma_{AB}|\gamma_{B})\geq -h_{2}(\epsilon,1-\epsilon)-\epsilon (S_{2}(\rho_{AB}|\rho_{B})-S_{2}(\tau_{AB}|\tau_{B}))\geq  -h_{2}(\epsilon,1-\epsilon)-2\epsilon \frac{d-1}{d} $. Second, as the conditional entropy is concave, $S_{2}(\gamma_{AB}|\gamma_{B})\geq (1-\epsilon)S_{2}(\rho_{AB}|\rho_{B})+\epsilon S_{2}(\tau_{AB}|\tau_{B})$, we obtain $S_{2}(\rho_{AB}|\rho_{B})-S_{2}(\gamma_{AB}|\gamma_{B})\leq \epsilon(S_{2}(\rho_{AB}|\rho_{B})-S_{2}(\tau_{AB}|\tau_{B}))\leq 2\epsilon \frac{d-1}{d}$, which completes the proof.

Using Lemma 6, and the method of \cite{Fannes}, we prove the result of Theorem 2.

\end{appendix}

\end{multicols}

\end{document}